\documentclass[aip,jcp,reprint,showkeys,superscriptaddress,a4paper,amsmath]{revtex4-1}
\usepackage{graphicx}
\usepackage{epsfig}
\usepackage{amsmath}
\usepackage{graphics}
\usepackage{latexsym}
\usepackage{amssymb}
\usepackage{inputenc}
\usepackage{xcolor}
\usepackage{siunitx}
\usepackage[colorlinks=true,linkcolor=magenta,citecolor=cyan]{hyperref}
\usepackage{threeparttable}

\newcommand{\la}{\left\langle}
\newcommand{\ra}{\right\rangle}

\begin{document}

\title{Novel stabilization mechanisms for concentrated emulsions with tunable morphology via amphiphilic polymer-grafted nanoparticles}

\author{Kojiro Suzuki}
\affiliation{Department of Mechanical Engineering, Keio University, 3-14-1, Hiyoshi, Kohoku-ku, Yokohama, 223-8522, Kanagawa, Japan}
\author{Yusei Kobayashi}
\affiliation{Department of Mechanical Engineering, Keio University, 3-14-1, Hiyoshi, Kohoku-ku, Yokohama, 223-8522, Kanagawa, Japan}
\affiliation{Faculty of Mechanical Engineering, Kyoto Institute of Technology, Goshokaido-cho, Matsugasaki, Sakyo-ku, Kyoto 606-8585, Japan}
\email{kobayashi@kit.ac.jp}
\author{Takashi Yamazaki}
\affiliation{KIRIN Central Research Institute, Kirin Holdings Company, Limited, 2-26-1 Muraoka-Higashi, Fujisawa, 251-8555, Kanagawa, Japan}
\author{Toshikazu Tsuji}
\affiliation{KIRIN Central Research Institute, Kirin Holdings Company, Limited, 2-26-1 Muraoka-Higashi, Fujisawa, 251-8555, Kanagawa, Japan}
\author{Noriyoshi Arai}
\affiliation{Department of Mechanical Engineering, Keio University, 3-14-1, Hiyoshi, Kohoku-ku, Yokohama, 223-8522, Kanagawa, Japan}

\begin{abstract}
This study explores the stabilization mechanisms of concentrated emulsions with tunable morphology using amphiphilic polymer-grafted nanoparticles (PGNPs). We employ coarse-grained molecular simulations to investigate concentrated oil-in-water emulsions stabilized by partially hydrolyzed poly(vinyl alcohol)-grafted poly(methyl methacrylate) (PMMA) particles. Two grafting architectures were examined: hydrophilic-hydrophobic (AB-type) diblock PGNPs and reverse BA-type diblock PGNPs. Our findings reveal that AB-type diblock PGNPs tend to aggregate, leading to droplet-droplet coalescence. In contrast, BA-type diblock PGNPs disperse effectively in the water phase, stabilizing robust emulsion through a space-filling mechanism. The study further demonstrates that the stability and morphology of the emulsions can be tuned by varying the number of PGNPs. Our results suggest that BA-type diblock PGNPs are more effective in stabilizing concentrated emulsions, offering insights for the design of novel emulsifiers in industrial applications.
\end{abstract}
\maketitle

\section{Introduction}
Concentrated emulsions with a dispersed phase of $>30\%$ are referred to as medium- or high-internal phase emulsions. Owing to their high volume fractions, they exhibit unique rheological and thermal properties that are not observed in conventional emulsions. This renders them as good candidates for potential applications in food, cosmetics, coatings, paints, pharmaceuticals, and porous material templates~~\cite{Ma:fh:2024,Zhao:japs:2022,Kim:pc:2022,Wang:cp:2024,Zong:fc:2022,Zhou:anm:2023,Zhou:aml:2023}. Large amounts of surfactants or amphiphilic polymers are frequently employed to prevent the coalescence of concentrated emulsions. However, careful selection is required because several of them exhibit cytotoxicity.

Concentrated emulsions stabilized by solid particles have recently attracted attention as a promising alternative because of their reduced toxicity and high attachment energy compared to conventional surfactants or polymers. In particular, the adsorption of solid particles at the oil/water interface acts as a physical barrier that impedes Ostwald ripening. These are also known as Pickering emulsions~\cite{Pickering:jcst:1907}. The stability of Pickering emulsions depends on the wettability of the solid particles~\cite{Everts:prl:2016,Bai:ass:2021,Wei:la:2023} and the interfacial assembly of their colloidal particles~\cite{Hammami:ami:2023,Alsmaeil:cm:2023,Inada:csa:2024}. Surface modification of particles is a versatile strategy for controlling the wettability and self-assembled structure at the liquid-liquid interface. Recent techniques have facilitated the synthesis of various polymers, resulting in various polymer-grafted nanoparticles (PGNPs) {\it via} chemical bonding or physical adsorption of polymers onto NP surfaces~\cite{Chancellor:ac:2019,Yuan:mrc:2021,Aldakheel:apm:2023}. A recent study showed that the partially-hydrolyzed poly (vinyl alcohol) (PVA) with an appropriate adjustment of the degree of saponification can be used as an effective emulsifier by combining experiments and molecular simulations~\cite{Suzuki:jml:2024}. Another study successfully prepared PVA-modified poly(methyl methacrylate) (PMMA) particles and evaluated the formulation of concentrated oil-in-water (O/W) Pickering emulsions stabilized by PMMA particles with $88\,{\rm mol\%}$ saponified PVA~\cite{yamazaki:csa:2024}. Because PVA is a water-soluble polymer with excellent biocompatibility and chemical properties and PMMA is a biocompatible non-stimuli-responsive polymer, concentrated O/W Pickering emulsions stabilized by PMMA–PVA have great potential applications in cosmetics, healthcare, emulsion-templated scaffolds, and biological applications such as tissue engineering.~\cite{yamazaki:csa:2024} However, the effects of the grafting architecture of PVA molecules on the interfacial assembly and stability of concentrated emulsions at the molecular level remain relatively unknown.

This study investigated the effects of the grafting architecture of partially hydrolyzed PVA-grafted PMMA NPs on the stability of concentrated O/W emulsions {\it via} coarse-grained molecular simulations. Two types of grafting architectures: (i) hydrophilic-hydrophobic (AB-type) diblock PGNPs (hydroxy groups as inner blocks and acetyl groups as outer blocks) and (ii) BA-type diblock PGNPs (opposite to the AB-type PGNPs) were considered. We discovered two different stabilization mechanisms that depend on the grafting architecture. In contrast to the Pickering stabilization by AB-type PGNPs, we observed that stable emulsion droplets were maintained although the BA-type PGNPs were not adsorbed on the droplets but were dispersed in the water phase. Furthermore, we found that BA-type PGNPs could significantly stabilize concentrated O/W emulsions over a wide range of volume fractions compared to AB-type PGNPs. Consequently, we demonstrated a novel pathway against droplet coalescence and coarsening and concluded that the formation of space-filling BA-type PGNPs between the emulsion droplets is the predominant stabilization mechanism of the concentrated emulsions.

\section{Methods and model}
\subsection{DPD method}
\label{sec:methods}
\begin{figure}[b!]
	\centering
	\includegraphics[width=8.5cm]{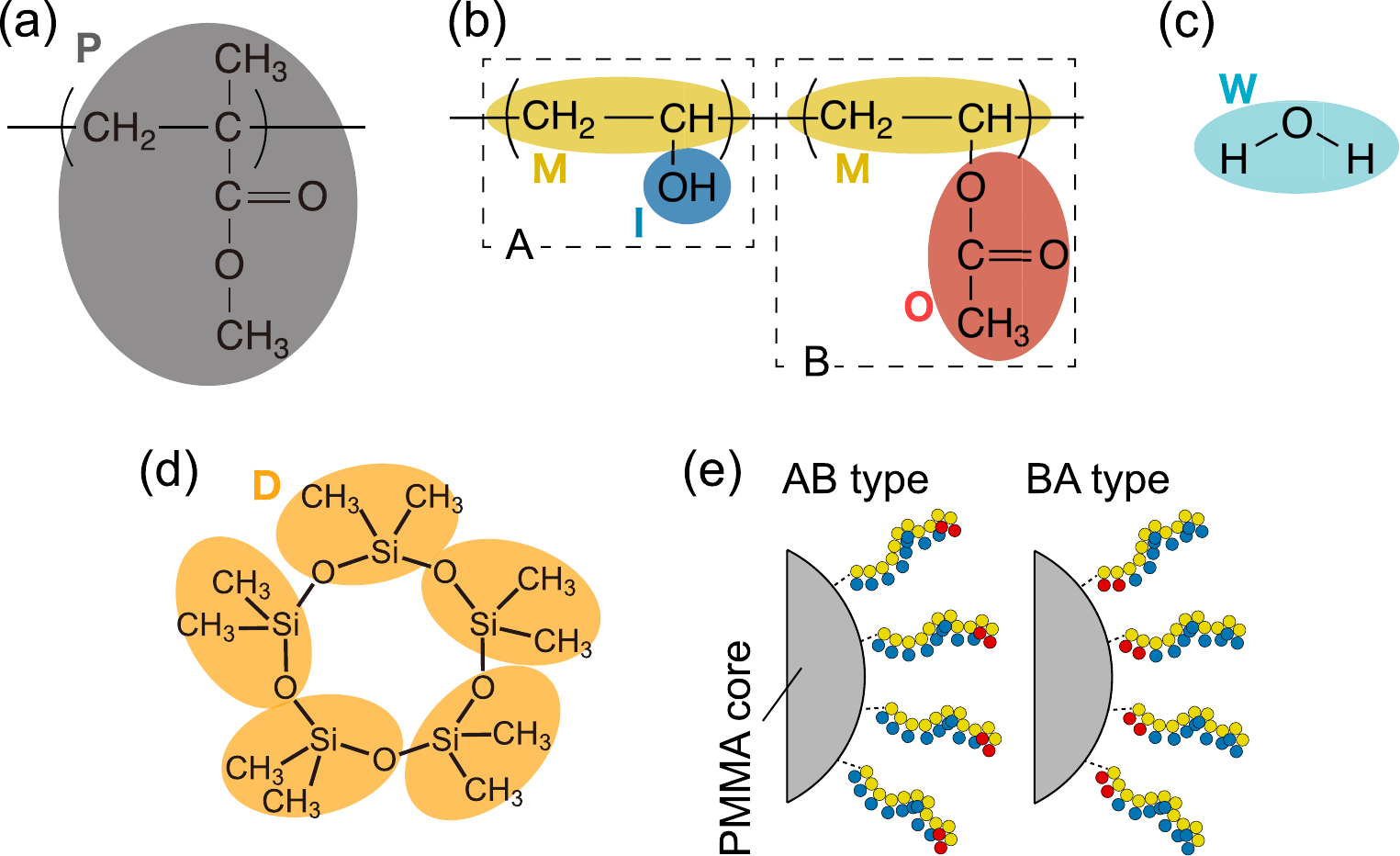}
	\caption{Molecular structures and coarse-grained models of (a) poly(methyl methacrylate) (PMMA), (b) poly(vinyl alcohol) (PVA), (c) water, and (d) Decamethylcyclopentasiloxane (DMCPS). Beads of a PMMA unit (gray); the vinyl (yellow), hydroxy (blue), and acetyl (red) groups of PVA; a DMCPS unit (orange); and a water unit (cyan), are denoted by P, M, I, O, D, and W respectively. (e) Schematic of two different polymer-grafted nanoparticles (PGNPs).}
	\label{fig:model}
\end{figure}
We employed the dissipative particle dynamics method~\cite{Hoogerbrugge:epl:1992,Groot:jpc:1997} which can simulate {\it millisecond} timescales and {\it micrometer} length scales by tracking the motion of coarse-grained particles (composed of a group of atoms or molecules). The DPD is based on Newton's equation of motion. Each coarse-grained particle (bead) $i$ in the system is subjected to three types of intermolecular forces: conservative, dissipative, and random. The equation of motion is as follows:
\begin{equation}
	m_i \frac{d {\bf{v}}_{i}}{dt} = {\bf{f}}_{i} = \sum_{j \neq i} {\bf{F}}_{ij}^\mathrm{C} + \sum_{j \neq i} {\bf{F}}_{ij}^\mathrm{D} + \sum_{j \neq i} {\bf{F}}_{ij}^\mathrm{R}\;\;,
	\label{eq:eq_motion}
\end{equation}
where $m$ is the mass, $\bf{v}$ is the velocity, $\bf{F}^{\rm{C}}$ is the conservative force, $\bf{F}^{\rm{D}}$ is the dissipative force, and $\bf{F}^{\rm{R}}$ is the pairwise random force. The sum of the forces acting on all beads between particles $i$ and $j$ is calculated. The conservative force is softly repulsive and is given by
\begin{equation}
	{\bf{F}}_{ij}^\mathrm{C} =
	\begin{cases}
		-a_{ij} \left( 1-\dfrac{ \left| {\bf{r}}_{ij}\right|}{r_{\mathrm c}} \right) {\bf{n}}_{ij}, & \left| {\bf{r}}_{ij} \right| \leq r_{\mathrm c}   \\
		\;\;\;\;\;\;\;\;\;\;\;\;\;\;\;0,                                                            & \left| {\bf{r}}_{ij} \right| > r_{\mathrm c}\;\;,
	\end{cases}
	\label{eq:FC}
\end{equation}
where ${\bf{r}}_{ij} = {\bf{r}}_{j} - {\bf{r}}_{i}$ and ${\bf{n}}_{ij} = {\bf{r}}_{ij} / \left| {\bf{r}}_{ij} \right|$. $a_{ij}$ describes the magnitude of the repulsive force between beads $i$ and $j$, and $r_{\mathrm c}$ is the cutoff distance. The dissipative force ${\bf{F}}_{ij}^\mathrm{D}$ and random force ${\bf{F}}_{ij}^\mathrm{R}$ can be expressed as
\begin{equation}
	\label{eq:FD}
	{\bf{F}}_{ij}^\mathrm{D} =
	\begin{cases}
		- \gamma \omega^{\mathrm D} \left( \left| {\bf{r}}_{ij} \right| \right) \left({\bf{n}}_{ij} \cdot {\bf{v}}_{ij} \right) {\bf{n}}_{ij}, & \left| {\bf{r}}_{ij} \right| \leq r_{\mathrm c} \\
		\;\;\;\;\;\;\;\;\;\;\;\;\;\;\;0,                                                                                                       & \left| {\bf{r}}_{ij} \right| > r_{\mathrm c}
	\end{cases}
\end{equation}
and
\begin{equation}
	\label{eq:FR}
	{\bf{F}}_{ij}^\mathrm{R} =
	\begin{cases}
		\sigma \omega^{\mathrm R} \left( \left| {\bf{r}}_{ij}\right| \right) \zeta_{ij} \Delta t^{-1/2} {\bf{n}}_{ij}, & \left| {\bf{r}}_{ij} \right| \leq r_{\mathrm c} \\
		\;\;\;\;\;\;\;\;\;\;\;\;\;\;\;0,                                                                               & \left| {\bf{r}}_{ij} \right| > r_{\mathrm c}
	\end{cases}
\end{equation}
where ${\bf{v}}_{ij} = {\bf{v}}_{j} - {\bf{v}}_{i}$, $\sigma$ is the noise parameter, $\gamma$ is the friction parameter, and $\zeta_{ij}$ is a random number based on a Gaussian distribution. Here, $\omega^{\mathrm R}$ and $\omega^{\mathrm D}$ are $r$-dependent weight functions expressed as
\begin{equation}
	\label{eq:w_func}
	\omega^{D} \left( r \right) = \left[ \omega^{R} \left( r \right) \right]^2 =
	\begin{cases}
		\left[1 - \dfrac{\left| {\bf{r}}_{ij} \right|}{r_{\rm c}}\right]^2, & r_{ij} \leq r_{\mathrm c}     \\
		\;\;\;\;0,                                                          & r_{ij} > r_{\mathrm c} \;\; .
	\end{cases}
\end{equation}
Temperature $T$ is controlled by the balance between ${\bf{F}}_{ij}^\mathrm{D}$ and ${\bf{F}}_{ij}^\mathrm{R}$. The values of $\sigma$ and $\gamma$ are related to each other by the fluctuation-dissipation theorem in the following equation:
\begin{equation}
	\label{eq:fd_theory}
	\sigma ^2 = 2 \gamma k_\mathrm{B} T,
\end{equation}
where $k_{\mathrm B}$ is the Boltzmann constant. Reduced units are often used in DPD simulations. Herein, the unit of length is the cutoff distance, $r_{\mathrm c}$, the unit of mass is the bead mass, $m$, and the unit of energy is $k_{\mathrm B}T$. Generally, temperature is expressed in the energy dimension ($k_{\mathrm B}T$).

\subsection{Models and conditions}
\begin{table}[!t]
\begin{center}
\begin{threeparttable}
\small
  \caption{Interaction parameters, $a_{ij}$, in ${\bf{F}}_{ij}^\mathrm{C}$ [see Eq.~\ref{eq:FC}] between all pairs. All values$^a$ are given in units of $k_{\rm B}T/r_{\rm c}$.}
  \label{aij_parameter}
  \begin{tabular*}{0.45\textwidth}{@{\extracolsep{\fill}}lllllll}
    \hline
     & P     & M     & I     & O     & D     & W\\
    \hline
    P& 18.75 & 18.95 & 67.48 & 19.51 & 19.78 & 35.06 \\
	M&       & 18.75 & 61.01 & 20.23 & 20.65 & 39.75 \\
    I&       &       & 18.75 & 48.77 & 48.10 & 23.94 \\
	O&       &       &       & 18.75 & 18.77 & 30.48 \\
	D&       &       &       &       & 18.75 & 29.70 \\
	W&       &       &       &       &       & 18.75 \\
  \end{tabular*}
\begin{tablenotes}
\item[a] P: PMMA, M: main chain (vinyl group) of PVA, I: hydrophilic side chain (hydroxy group) of PVA, O: hydrophobic side chain (acetyl group) of PVA. D: DMCPS, W: Water
\end{tablenotes}
\end{threeparttable}
\end{center}
\end{table}
The coarse-grained PVA-grafted PMMA NP models were basically constructed based on previous studies~\cite{Suzuki:jml:2024,Inada:csa:2024,yamazaki:csa:2024}. Each core of PGNP comprises 162 DPD beads and its radius, $R_{\rm NP}$, is set to $2.0\,r_{\rm c}$. The vertex beads of the PMMA NP (denoted by P in Fig.~\ref{fig:model}(a)) are placed on the vertices of a regular icosahedron and connected to their nearest neighbors and the diametrically opposite bead through a harmonic potential $U_{\rm b}$:
\begin{equation}
	U_{\rm b}(r_{ij}) = \frac{k}{2}(r_{ij}-r_{0})^2,
	\label{eq:Ubond}
\end{equation}
where $k$ is the spring constant, $r_{ij}$ is the distance between beads $i$ and $j$, and $r_{0}$ is the desired distance between two bonded vertex beads. We set $k=5000\,k_{\rm B}T/r_{\rm c}^2$~\cite{Kobayashi:la:2020,Poblete:pre:2014,Kobayashi:sm:2020,wani:jcp:2022,wani:sm:2024} to maintain a (nearly) rigid NP shape with Boltzmann constant $k_{\rm B}$ and temperature $T$. The molecular structures and coarse-grained models used in this study are shown in Figs.~\ref{fig:model}(a-d). The PVA model comprised three types of beads: the main chain (vinyl group), hydrophilic side chain (hydroxy group), and hydrophobic side chain (acetyl group) (denoted as M [yellow region], I [blue region], and O [red region], respectively, in Fig.~\ref{fig:model}(b)). A previous study showed that stable, concentrated emulsions were obtained with a smaller amount of PVA when the degree of saponification was tuned to a suitable value (80\%). Thus, the degree of saponification $f=a/(a+b)$ was fixed at $f=0.8$, where $a$ and $b$ are the number of hydrophilic and hydrophobic groups, respectively. As shown in Fig.~\ref{fig:model}(c) and (d), decamethylcyclopentasiloxane (DMCPS) and solvent (water) molecules are represented by a single bead and denoted by O and W, respectively. The nearest-neighbor particles in the PVA and DMCPS molecules are also connected through $U_{\rm b}$ with $k=100\,k_{\rm B}T/r_{\rm c}^2$ and equilibrium bond length $r_{0}=0.5\,r_{\rm c}$. In addition, a bending potential was included between two adjacent bonds in DMCPS with $k_{\rm angle}=12\,k_{\rm B}/{\rm rad}^2$ and $\theta_{0}=108^\circ$. The effect of the grafting architectures was investigated by considering two types of PGNP: (i) hydrophilic-hydrophobic (AB-type) diblock PGNPs (hydrophilic monomers as inner blocks and hydrophobic monomers as outer blocks) and (ii) BA-type diblock PGNPs (opposite to AB-type PGNPs). To obtain visually interpretable representations, schematics of the two different PGNPs are shown in Fig.~\ref{fig:model}(e). The graft density was set as $\sigma_{\rm g}= 3.2$; therefore, the number of grafted chains was set as 162.

The parameters describing the interaction between beads of the same type, $a_{ii}$, are related to the density of the beads in the system, $\rho$, and the degree of coarse graining based on liquid compressibility~\cite{Groot:jcp:1998}. In this study, $a_{ii}$ was set to 18.75\,$k_{\rm B}T$ using Eq.~\ref{eq:aii}:
\begin{equation}
	a_{ii} = 75\,k_{\rm B}T/\rho.
	\label{eq:aii}
\end{equation}
The parameters describing the interaction between the different types of beads $a_{ij}$ are defined as follows:
\begin{equation}
	a_{ij} = a_{ii} + 3.268\frac{V_{\rm seg}}{RT}(\delta_i - \delta_j)^2
	\label{eq:aij}
\end{equation}
where $V_{\rm seg}$ is the average molar volume of beads $i$ and $j$, $R$ is the gas constant, and $\delta$ is the bead solubility parameter. The interaction parameters between any two DPD beads were inspired by recent studies~\cite{Suzuki:jml:2024,yamazaki:csa:2024} and were estimated using J-OCTA simulation software~\cite{jocta} and the Fragment Molecular Orbital (FMO)-based $\chi$-parameter Evaluation Workflow System (FCEWS) package~\cite{okuwaki:jpcb:2018,okuwaki:jccj:2018}. The obtained values are presented in Table~\ref{aij_parameter}. The complete details of this procedure are provided in Refs.~~\cite{Suzuki:jml:2024,yamazaki:csa:2024}.

Based on a previous study~\cite{Suzuki:jml:2024}, four oil droplets, each with a face-centered cubic structure, were immersed in the water phase in the initial configuration. Several PGNPs were randomly placed at the oil-water interface. We varied the number of PGNPs per oil droplet as $N_{\rm NP}=30-60$, which resulted in $120-240$ PGNPs in the system. The ratio of the number of beads in the oil phase to that in the water phase was set to 6:4\,(wt.\%); thus, the numbers of beads in the oil and water phases were 64,800 and 43,200, respectively. The length of the edge of the cubic simulation box was adjusted to achieve $\rho=4\,r_{\rm c}^{-3}$ and the radius of the oil droplets was $\sim 10\,r_{\rm c}$. Periodic boundary conditions were applied to all the three dimensions. The temperature was set as $1.5\,k_{\rm B}T$. The noise parameter $\sigma$ was set to 3.0, the friction parameter $\gamma$ was set to 4.5, and the time step $dt$ was 0.01. All simulations were performed using the HOOMD-Blue software package (version~2.9.7)~\cite{anderson:cms:2020,Phillips:jcp:2011}.

\section{Results and discussion}
\label{sec:results}
\subsection{Mechanisms of stabilization}
\begin{figure}[tb!]
	\centering
	\includegraphics[width=8.5cm]{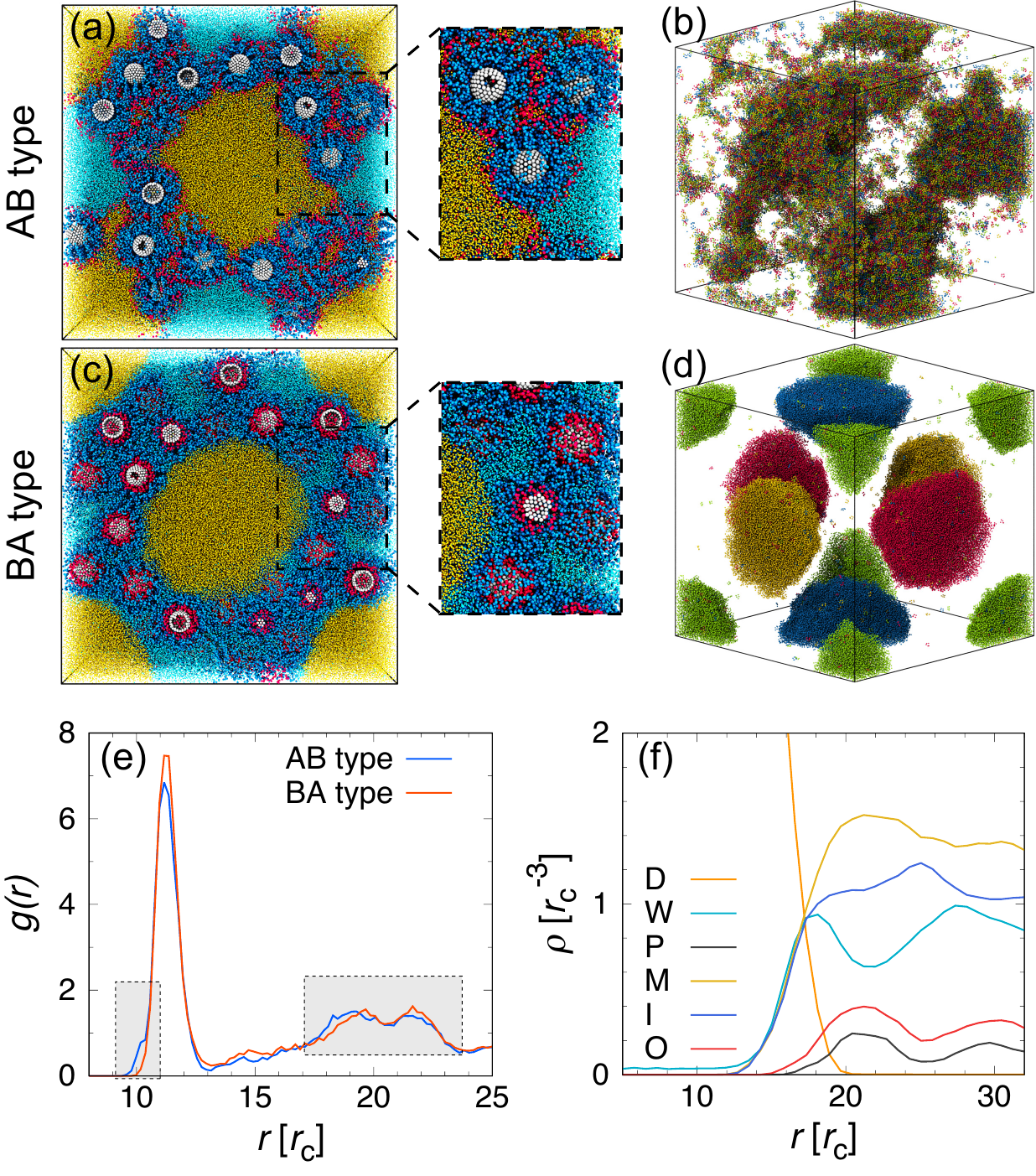}
	\caption{Comparison of the emulsified state for (a and b) AB-type PGNPs and (c and d) BA-type PGNPs at $N_{\rm NP}=50$, respectively. (b and d) For clarity, the oil beads are only shown, and the droplets in the representative snapshots are colored for optimal visualization. (e) Radial distribution function, $g(r)$, between the center of mass of the NPs. (f) Radial density distributions from the center of mass of the oil droplets at $N_{\rm NP}=50$ for BA-type PGNPs. D: Decamethylcyclopentasiloxane, W: Water, P: poly(methyl methacrylate), M: main chain (vinyl group) of PVA, I: hydrophilic side chain (hydroxy group) of PVA, O: hydrophobic side chain (acetyl group) of PVA.}
	\label{fig:combined_dens_gr}
\end{figure}

We began by comparing the emulsified states of the O/W concentrated emulsions for two different grafting architectures. Figures~\ref{fig:combined_dens_gr}(a-d) show representative snapshots of the AB- and BA-type PGNPs with $N_{\rm NP}=50$. In Figs.~\ref{fig:combined_dens_gr}(b) and (d), only the oil beads are shown, and each bead in the snapshots is colored based on the group of four oil droplets in the initial configurations. Significant differences were observed between the emulsified states of the two types of PGNPs. The AB-type PGNPs were adsorbed onto the droplets because of the interaction between the outer hydrophobic block of the grafted PVA and the oil droplets (Fig.~\ref{fig:combined_dens_gr}(a)). Nevertheless, partial droplet-droplet coalescence occurred, and the DMCPS particles mixed with those of the other oil droplets (Fig.~\ref{fig:combined_dens_gr}(b)). This result is somewhat surprising because the adsorption of PGNPs at the oil-water interface is crucial for avoiding the coalescence of emulsion droplets. However, in the case of the O/W emulsion, the outer hydrophobic blocks of the PGNPs caused not only adsorption onto the oil-water interface but also aggregation of the PGNPs (magnified view of Fig.~\ref{fig:combined_dens_gr}(a)). For quantitative analysis, we provide the radial distribution function $g(r)$ between the centers of mass of the NPs in Fig.~\ref{fig:combined_dens_gr}(e). We observed a slight increase in the proportion of the first, second, and third peaks in the $g(r)$ curve between the AB-type PGNPs before that in the $g(r)$ curve between the BA-type PGNPs. Thus, a tighter aggregation was observed between AB-type PGNPs and BA-type PGNPs. This also explains why the shape of the emulsion droplets covered by the AB-type PGNPs was deformed from a sphere, compared to the case of the BA-type PGNPs ({\it cf.} Fig.~\ref{fig:combined_dens_gr}(a) and (c)). Thus, for the O/W emulsion, the outer hydrophobic block of grafted PVA molecules became the ``sticky'' points for the adhesion of both emulsion droplets and each PGNP. This resulted in droplet-droplet coalescence even for effective emulsifiers in reducing interfacial tension~\cite{Inada:csa:2024}.

\begin{figure}[htb!]
	\centering
	\includegraphics[width=8.0cm]{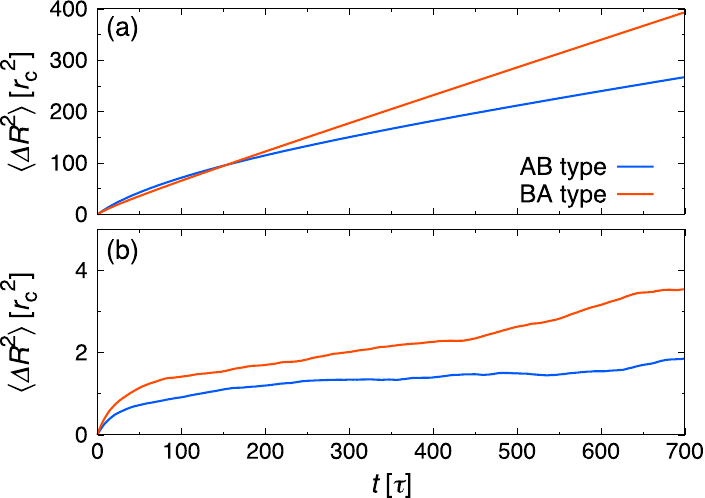}
	\caption{Mean square displacement, $\la \Delta R^2 \ra$, of (a) water and (b) center of mass of nanoparticles with $N_{\rm NP}=50$.}
	\label{fig:msd}
\end{figure}
\begin{figure*}[htb!]
	\centering
	\includegraphics[width=12.5cm]{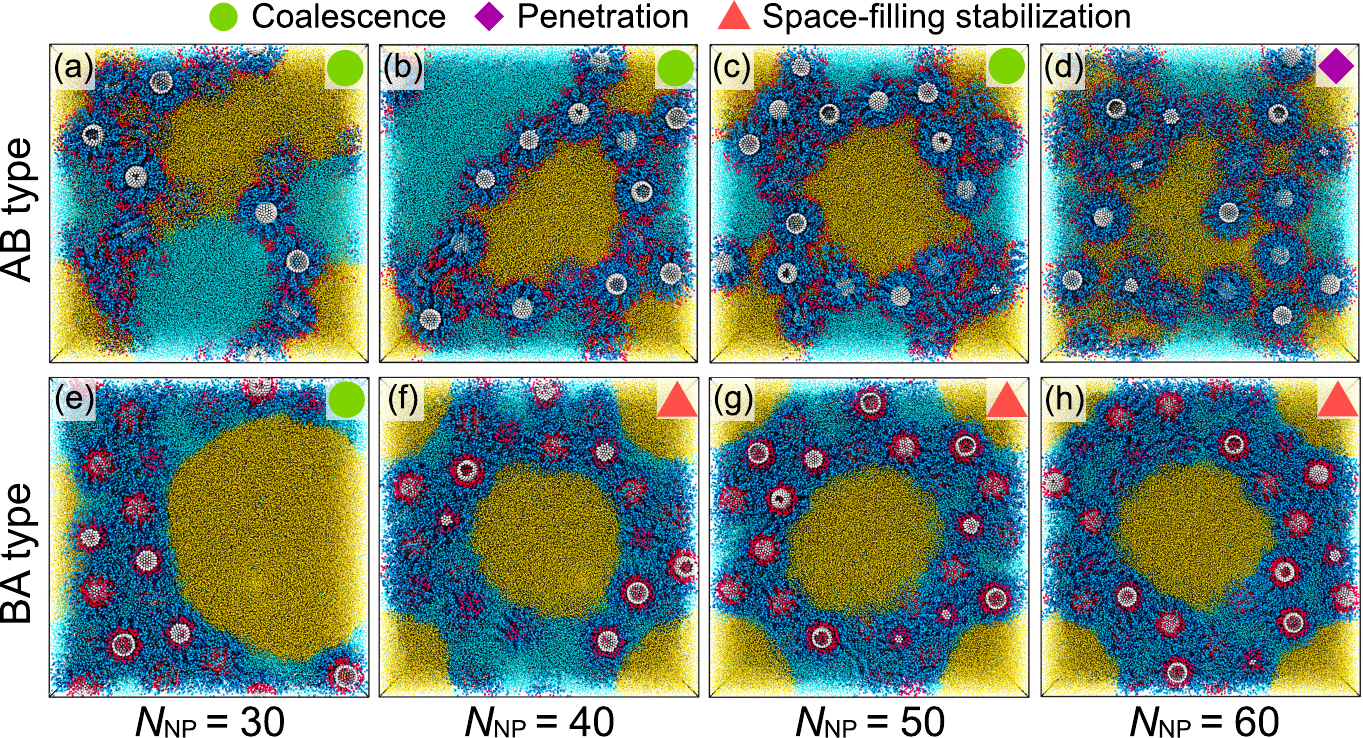}
	\caption{$N_{\rm NP}$ dependence of the morphologies and emulsified states for (a-d) AB-type PGNPs and (e-h) BA-type PGNPs.}
	\label{fig:phi}
\end{figure*}
Interestingly, in contrast to the AB-type PGNPs, stable emulsion droplets were maintained although the BA-type PGNPs were not adsorbed on the droplets and were rather dispersed in the water phase, as shown in Fig.~\ref{fig:combined_dens_gr}(c) and (d). For a more quantitative explanation, we analyzed the radial density distributions from the center of mass of the oil droplets (Fig.~\ref{fig:combined_dens_gr}(f)). We confirmed (nearly) zero density ($\rho\approx 0\,r_{\rm c}^{-3}$) at $r\gtrsim 20\,r_{\rm c}$, indicating that stable emulsion droplets were maintained. For the molecular distributions around the oil droplets, the density peak of water appeared before the peaks of other species. This indicated the existence of a mechanism governing effective emulsion stabilization that differed from the adsorption of solid particles at the droplet surface (Pickering stabilization). To understand the mechanisms underlying these results, we considered molecular diffusion, which is an important factor in determining emulsion stability. It might be possible to observe differences in the diffusion of the molecules between two different emulsion stabilizations, {\it that is}, Pickering stabilization by AB-type PGNPs and space-filling stabilization by BA-type PGNPs ({\it cf.} Fig.~\ref{fig:combined_dens_gr}(a) and (c)). We computed the mean squared displacement of the water and NP centers of mass, $\Delta R^2(t) = \la \left[\mathbf{R}_i(t_0+t) - \mathbf{R}_i(t_0)\right]^2 \ra$, as a function of time; the data are shown in Fig.~\ref{fig:msd}(a) and (b). Regarding the diffusion behavior of water, the slope for the AB-type PGNPs was smaller than that for the BA-type PGNPs (Fig.~\ref{fig:msd}(a)). This is primarily because this difference can be attributed to the ability of water particles in the system with the BA-type PGNPs to effectively disperse between the oil droplets. The BA-type PGNPs did not aggregate with each other; rather, their outer hydrophilic blocks preferred to be in contact with water, resulting in a uniform distribution of water between the oil droplets. In contrast, the AB-type PGNPs formed self-assembled aggregates owing to the hydrophobic effect of the outer blocks. In addition, the aggregates exposed to water did not prefer to be in contact with water, thus preventing the uniform distribution of water between the oil droplets. Compared to the well-dispersed distribution, the diffusivity of the localized water decreases because of the formation of aggregates of PGNPs. We examined the diffusion behavior of PGNPs, as shown in Fig.~\ref{fig:msd}(b). The same trend was observed as seen in the water case; thus, the slope in the case of the BA-type PGNPs was greater than that for the AB-type PGNPs. This result indicates the evidence of effective emulsion stabilization that differed from the adsorption of solid particles at the droplet surface (Pickering stabilization). Thus, the completely space-filling PGNPs in the interdroplet spaces between the emulsion droplets effectively act as a barrier to droplet-droplet coalescence. We demonstrated a novel pathway against droplet coalescence and coarsening and concluded that the formation of space-filling BA-type PGNPs between emulsion droplets is the predominant stabilization mechanism of concentrated emulsions.

\subsection{Tuning morphology and emulsion stability}
We systematically varied the number of PGNPs, $N_{\rm NP}$, to further investigate the morphology and stability of the O/W concentrated emulsions using the two different diblock PGNPs. Figures~\ref{fig:phi}(a-h) show the $N_{\rm NP}$ dependence of the emulsified state for different types of PGNPs. We observed three characteristic states: droplet coalescence (circles), penetration of PGNPs into oil droplets (diamonds), and stable droplets by space-filling PGNPs stabilizing (triangles). For $N_{\rm NP}\le 30$, no stable concentrated O/W emulsion was observed, and droplet-to-droplet coalescence occurred for both types of PGNPs (Fig.~\ref{fig:phi}(a) and (e)), reflecting insufficient surface coverage. When $N_{\rm NP}$ was increased further, a difference in the morphology and stability of the emulsions was observed between the AB and BA-type PGNPs. For $40 \le N_{\rm NP}\le 50$, droplet-to-droplet coalescence still occurred in the system containing the AB-type PGNPs (Fig.~\ref{fig:phi}(b) and (c)), whereas stable concentrated O/W emulsions were formed only in the system containing the BA-type PGNPs (Fig.~\ref{fig:phi}(f) and (g)). These results indicate that the BA-type PGNPs, which enable space-filling stabilization, were more effective emulsifiers than the AB-type PGNPs. At the highest investigated number of PGNPs ($N_{\rm NP}=60$), the emulsified state was maintained for BA-type PGNPs, as shown in Fig.~\ref{fig:phi}(h). The fact that BA-type PGNPs can significantly stabilize concentrated O/W emulsions over a wide range of $N_{\rm NP}$ values supports the finding that BA-type PGNPs are more effective than AB-type PGNPs. In contrast, even at high $N_{\rm NP}$ for the AB-type PGNPs, no stable concentrated O/W emulsion was observed; rather, the penetration of PGNPs into the oil droplets was observed. The PGNPs were no longer adsorbed onto the droplets; therefore, Pickering stabilization can only be realized over a narrow range of $N_{\rm NP}$.

\section{Conclusions}
\label{sec:conclusions}
This study performed coarse-grained molecular simulations to investigate the effects of the grafting architecture of partially hydrolyzed PVA-grafted PMMA NPs on the emulsion stability of concentrated oil-in-water (O/W) emulsions. Two different stabilization mechanisms with tunable morphology were identified depending on the grafting architectures. When the hydroxy group was the inner block and acetyl group was the outer block, the PGNPs were adsorbed onto the droplet. This facilitated Pickering stabilization owing to the interaction between the outer hydrophobic block of the grafted PVA and the oil droplets. However, the outer hydrophobic block of grafted PVA molecules became the ``sticky'' points for the adhesion of both emulsion droplets and each PGNP, resulting in droplet-droplet coalescence. When the acetyl group was the inner block and hydroxy group was the outer block, the PGNPs were not adsorbed on the droplets; rather, they were dispersed in the water phase. Nevertheless, stable emulsion droplets were maintained over a wide range of volume fractions. Further, we demonstrated a novel pathway against droplet coalescence and coarsening, which differed from the adsorption of solid particles on the droplet surface. Thus, the formation of space-filling PGNPs between the emulsion droplets was the predominant stabilization mechanism for the concentrated emulsions. Our findings indicate that careful tuning of the grafting architecture is necessary to control the stability of concentrated O/W emulsions using amphiphilic diblock PVA-grafted PMMA NPs. Our simulations offer a theoretical guide for controlling the morphologies and stabilities of concentrated O/W emulsions via different grafting architectures, paving the way for novel emulsification strategies that may find applications in food, cosmetics, coatings, paints, pharmaceuticals, and porous material templates.

%

\section*{Declaration of Competing Interest}
The authors declare the following financial interests/personal relationships which may be considered as potential competing interests: Takashi Yamazaki reports a relationship with Kirin Holdings Company Limited that includes: employment. Toshikazu Tsuji reports a relationship with Kirin Holdings Company Limited that includes: employment. If there are other authors, they declare that they have no known competing financial interests or personal relationships that could have appeared to influence the work reported in this paper.

\section*{Acknowledgements}
Part of this work was carried out using the supercomputer Fugaku (Project numbers ~hp230016 and ~hp240013). Y.K. acknowledges JSPS KAKENHI Grant No.~JP24K17216 and the support of KIT Grants-in-Aid for Early-Career Scientists.


\bibliography{cite} 

\end{document}